\documentclass{PoS}

\title{GFMC calculations of electromagnetic moments and M1 transitions in $A\leq 9$ nuclei}

\ShortTitle{GFMC calculations of electromagnetic moments and M1 transitions in A $\leq$ 9 nuclei}

\author{\speaker{S.\ Pastore}\\
        Department of Physics and Astronomy, University of South Carolina, Columbia, SC 29208\\
        E-mail: \email{pastores@mailbox.sc.edu}}

\author{Steven C. Pieper\\
        Physics Division, Argonne National Laboratory, Argonne, Illinois 60439\\
        E-mail: \email{spieper@anl.gov}}

\author{R. Schiavilla\\
        Theory Center, Jefferson Laboratory, Newport News, Virginia 23606,\\
        Department of Physics, Old Dominion University, Norfolk, Virginia 23529\\
        E-mail: \email{schiavil@jlab.org}}

\author{R. B.\ Wiringa\\
        Physics Division, Argonne National Laboratory, Argonne, Illinois 60439\\
        E-mail: \email{wiringa@anl.gov}}

\abstract{We present recent Green's function Monte Carlo calculations
of magnetic moments and M1 transitions in $A \leq 9$ nuclei, which include
corrections arising from two-body meson-exchange electromagnetic currents. 
Two-body effects provide significant corrections to the calculated observables,
bringing them in excellent agreement with the experimental data.
In particular, we find that two-body corrections are especially large in
the $A=9$, $T=3/2$ systems, in which they account for up to $\sim 20\%$ ($\sim 40 \%$)
of the total predicted value for the $^9$Li ($^9$C) magnetic moment.
}
\FullConference{The 7th International Workshop on Chiral Dynamics,\\
		August 6 - 10, 2012\\
		Jefferson Lab, Newport News, Virginia, USA}
\begin{document}

\begin{figure}
\begin{center}
\includegraphics[height=.237\textheight]{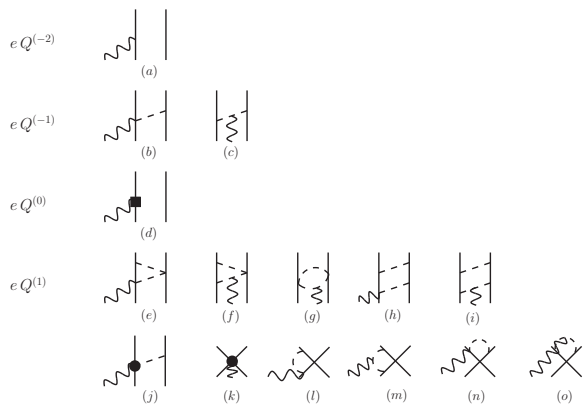}
\caption{Diagrams illustrating one- and two-body $\chi$EFT EM currents entering at LO ($e\, Q^{-2}$),
NLO ($e\, Q^{-1}$), N2LO  ($e\, Q^{\,0}$), and N3LO ($e\, Q^{\,1}$).  Nucleons, pions,
and photons are denoted by solid, dashed, and wavy lines, respectively.}
\label{fig:f1}
\end{center}
\end{figure}
In this contribution, we present a set of Green's function Monte Carlo (GFMC) calculations
of magnetic moments (m.m.'s) and M1 transitions in $A\leq9$ nuclei, which has been recently
reported in Ref.~\cite{Pastore12}. In these calculations, nuclear wave functions (w.f.'s)
are constructed from a Hamiltonian consisting of the Argonne-$v_{18}$ two-nucleon~\cite{WSS95}
and Illinois-7 three-nucleon potentials~\cite{P08b}, with which the computed GFMC ground- and
excited-state energies are found to be in good agreement with experiments~\cite{Pastore12}.
The electromagnetic (EM) current operator includes, in addition to the standard one-body
convection and spin-magnetization terms for individual protons and neutrons, a two-body
meson-exchange-current (MEC) component.
The latter is constructed within two distinct frameworks, namely the standard nuclear
physics approach (SNPA) illustrated in Refs.~\cite{Mar05,MPPSW08},
and the pionfull chiral effective field theory ($\chi$EFT) formulation of
Refs.~\cite{Pastore08,PGSVW09,Piarulli12}. In what follows, we summarize on the methods
and results discussed in Ref.~\cite{Pastore12}.

\section{GFMC Method and the Nuclear Hamiltonian}

The EM transition matrix elements are evaluated in between w.f.'s which are solutions
of the Schr\"odinger equation, with a nuclear Hamiltonian, H, consisting of a kinetic term plus
two- and three-body interaction terms---in the present case, the Argonne-$v_{18}$ and Illinois-7, respectively.
Nuclear w.f.'s are constructed in two steps. First, a trial variational Monte Carlo w.f. ($\Psi_T$),
which accounts for the effect of the nuclear interaction via the inclusion of correlation operators,
is generated by minimizing the energy expectation value with respect to a number of
variational parameters. The second step improves on $\Psi_T$ by eliminating
excited states contamination. This is accomplished by the GFMC
calculation which propagates the Schr\"odinger equation in imaginary time ($\tau$). The propagated w.f.
$\Psi(\tau) = e^{-(H-E_0)\tau}\Psi_T$, for large values of $\tau$, converges to the
exact w.f. with eigenvalue $E_0$.  Ideally, the matrix elements should be evaluated in between
two propagated w.f.'s. In practice, we evaluate mixed estimates in which only one w.f. is propagated,
while the remaining one is replaced by $\Psi_T$. The calculation of diagonal and
off-diagonal matrix elements is discussed at length in Ref.~\cite{PPW07} and references therein. 

The nuclear EM current operator is also expressed as an expansion in many-body operators.
The current utilized in the calculations accounts up to two-body effects, and is written as:
\begin{equation}
 {\bf j}({\bf q}) =  \sum_i {\bf j}_i({\bf q}) + \sum_{i<j} {\bf j}_{ij}({\bf q}) \ ,
\end{equation}
where ${\bf q}$ is the momentum associated with the external EM field. The one-body operator
at leading order, {\it i.e.} the impulse approximation (IA) operator, consists of the convection and
the spin-magnetization currents associated with an individual nucleon~\cite{Pastore12}, and
it is diagrammatically represented in panel (a) of Fig.~\ref{fig:f1}. 

\section{$\chi$EFT and SNPA EM currents}

In the calculations, two models for the EM two-body MEC operators are tested,
namely the pionful $\chi$EFT and SNPA models.
The $\chi$EFT current operators are expanded in powers of pions' and nucleons'
momenta, $Q$, and consist of long- and intermediate-range components which are described in
terms of one- and two-pion exchange contributions, as well as contact currents which
encode the short-range physics. These operators involve a number of Low Energy Constants (LECs)
which are then fixed to the experimental data.  Currents from pionful $\chi$EFT including up
to two-pion exchange contributions were derived originally by Park, Min, and Rho in
covariant perturbation theory~\cite{Park96}. More recently, K\"olling and collaborators
presented EM currents obtained within the method of unitary transformations~\cite{Kolling09,Kolling11}.
Here, we refer to the EM operators constructed in Ref.~\cite{Pastore08,PGSVW09,Piarulli12}, in which
time-ordered perturbation theory is implemented to calculate the EM transition amplitudes.
These EM operators are diagrammatically represented in Fig.~\ref{fig:f1}, where they are listed
according to their scaling in $e Q$, (where $e$ is the electric charge).

Referring to this figure, one-body contributions enter at LO, panel (a),
and N2LO, panel (d), and they are the IA current operator at LO and its relativistic
correction, respectively. The NLO term involves seagull and in-flight
long-range contributions associated with one-pion exchange (OPE).
At N3LO we include the two-pion-range contributions of diagrams (e)--(i),
the one-pion-range tree-level current involving a $\gamma \pi NN$ vertex of order $e\, Q^2$, diagram (j),
the contact currents of diagram (k), as well as the one-loop corrections of diagrams
(l)--(o). The two-body operators have a power-law behavior at large momenta, therefore
a regularization procedure is implemented via the introduction of cutoff function of the form 
$exp(-Q^4/\Lambda^4)$~\cite{Piarulli12}, where $\Lambda=600$ MeV.

The contact currents of diagram (k) are of minimal and non-minimal
nature. The former are linked to the $\chi$EFT potential at order $Q^2$
via current conservation; therefore they involve the same LECs entering the
$\chi$EFT NN interaction, and can be taken from fits to the NN scattering
data. We use the values obtained from the analysis of Refs.~\cite{Entem03},
with cutoff $\Lambda=600$ MeV.    
Non-minimal LECs entering the contact and tree-level currents at N3LO---diagrams (j)
and (k), respectively---need to be fixed to EM observables. 
The fitting procedure has been implemented by Piarulli {\it et al.} in Ref.~\cite{Piarulli12}.
In that work, LECs multiplying isovector operators in the tree-level current are
fixed by saturating the $\Delta$-resonance~\cite{Park96} (a common strategy adopted, for example, in
Refs.~\cite{Song07,Song09,Lazauskas11}). The remaining three LECs are
fixed so as to reproduce the deuteron, $^3$He, and $^3$H m.m.'s.

The second model for the EM MEC operators utilized in the calculations
is the SNPA model. Two-body currents in the SNPA formalism, described at length in
Refs.~\cite{Mar05,MPPSW08} and references therein, are separated into model-independent (MI) and model-dependent
(MD) terms. The former (MI) are derived from the $N\!N$ potential, and their longitudinal
components satisfy, by construction, current conservation with it, thus their
short-range behavior is consistent with that of the potential. The dominant terms,
isovector in character, originate from the static part of the potential, which is assumed
to be due to exchanges of effective pseudoscalar (PS or ``$\pi$-like'') and vector
(PV or ``$\rho$-like'') mesons.  The associated currents are then constructed by
using the PS and PV propagators, projected out of the static potential~\cite{Mar05}.
At large inter-nucleon separations, where the $N\!N$ potential is driven by the OPE
mechanism, the MI current coincides with the standard seagull and pion-in-flight
OPE currents diagrammatically illustrated in panels (b) and (c), respectively, of
Fig.~\ref{fig:f1}. The MD currents are purely transverse, and unconstrained by current
conservation.  The dominant term is associated with excitation of
intermediate $\Delta$ isobars. Additional and small MD currents arise from the
isoscalar $\rho\pi\gamma$ and isovector $\omega\pi\gamma$ transition mechanisms~\cite{Mar05,MPPSW08}. 

\begin{figure}
\begin{minipage}[t]{0.32\textheight}
\includegraphics[height=.310\textheight,angle=270,keepaspectratio=true]{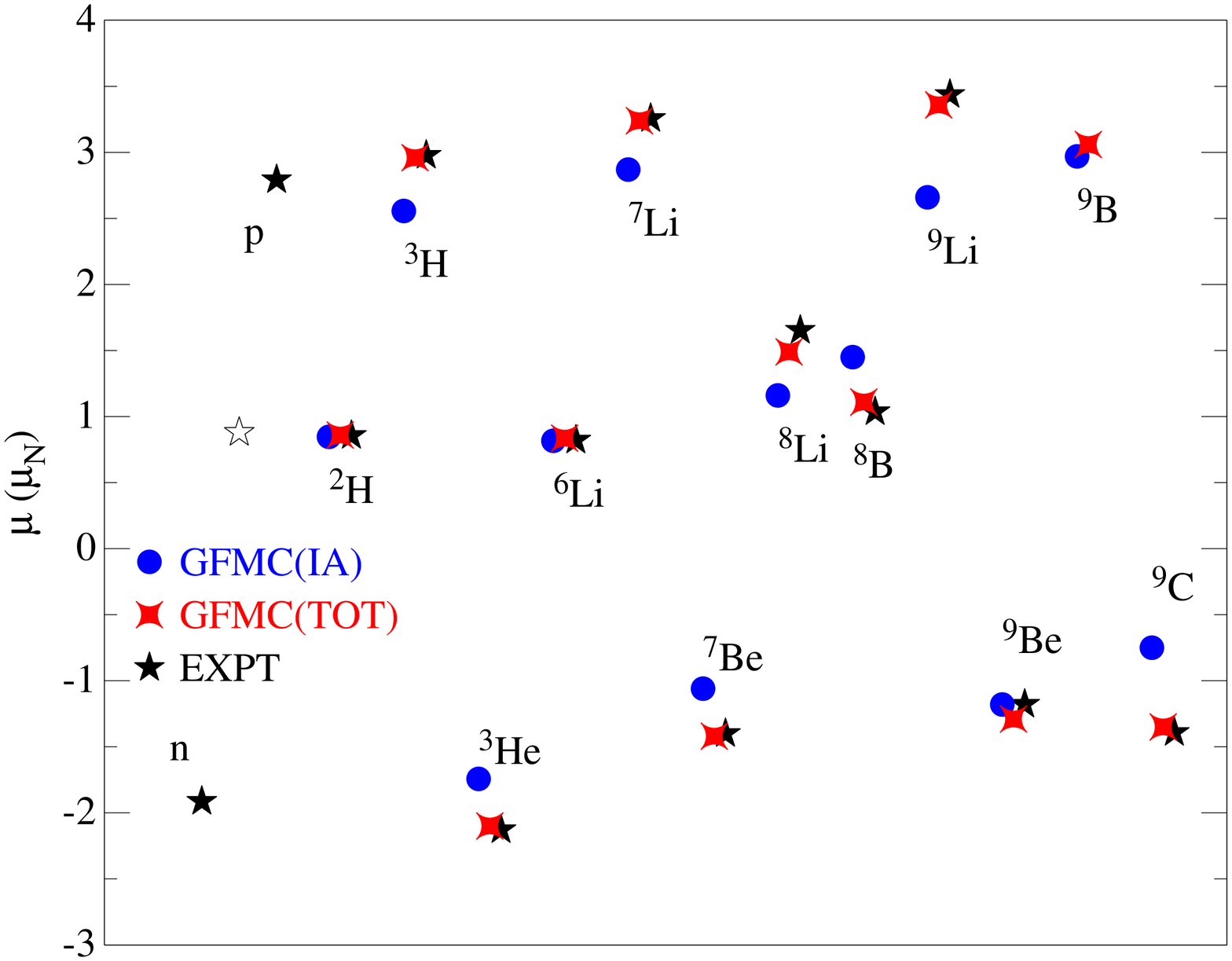}
\end{minipage}
\begin{minipage}[t]{0.30\textheight}
\vspace{0.08in}
\includegraphics[height=.35\textheight,angle=270,keepaspectratio=true]{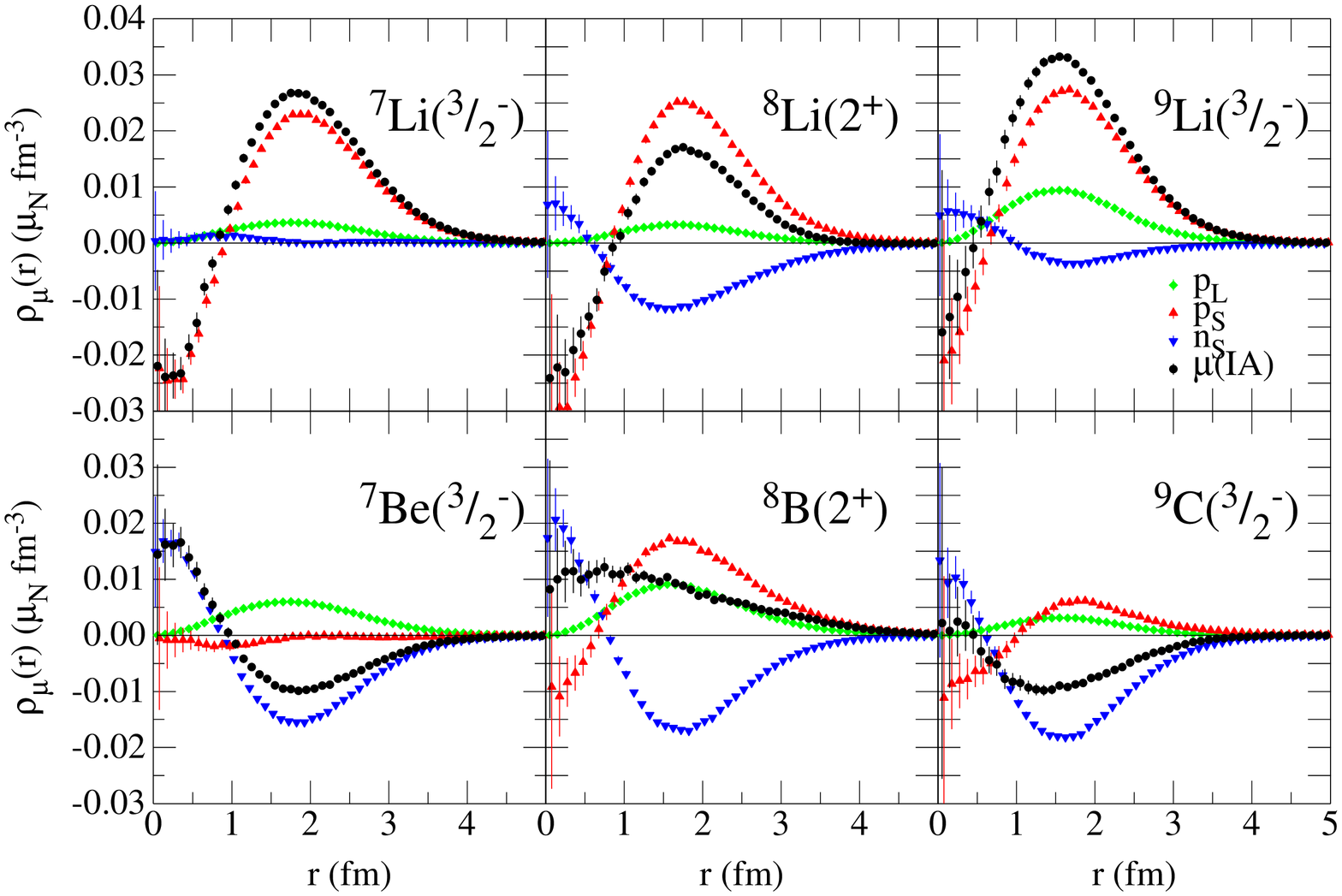}
\label{fig:f2}
\end{minipage}
\caption{  {\bf Left}: Magnetic moments in nuclear magnetons for $A\leq9$ nuclei. 
           Black stars indicate
           the experimental values~\cite{Tilley02,Tilley04}, while blue dots (red diamonds)
           represent preliminary GFMC calculations which include the IA one-body EM current
           (full $\chi$EFT current up to N3LO). Predictions are for nuclei with $A>3$. {\bf Right}:
           Magnetic density in nuclear magnetons per fm$^3$ for selected nuclei, including only
           the IA current contribution.
           }
\end{figure}

\section{Results}

\begin{figure}
\begin{minipage}[t]{0.30\textheight}
\begin{center}
\includegraphics[height=.310\textheight,keepaspectratio=true]{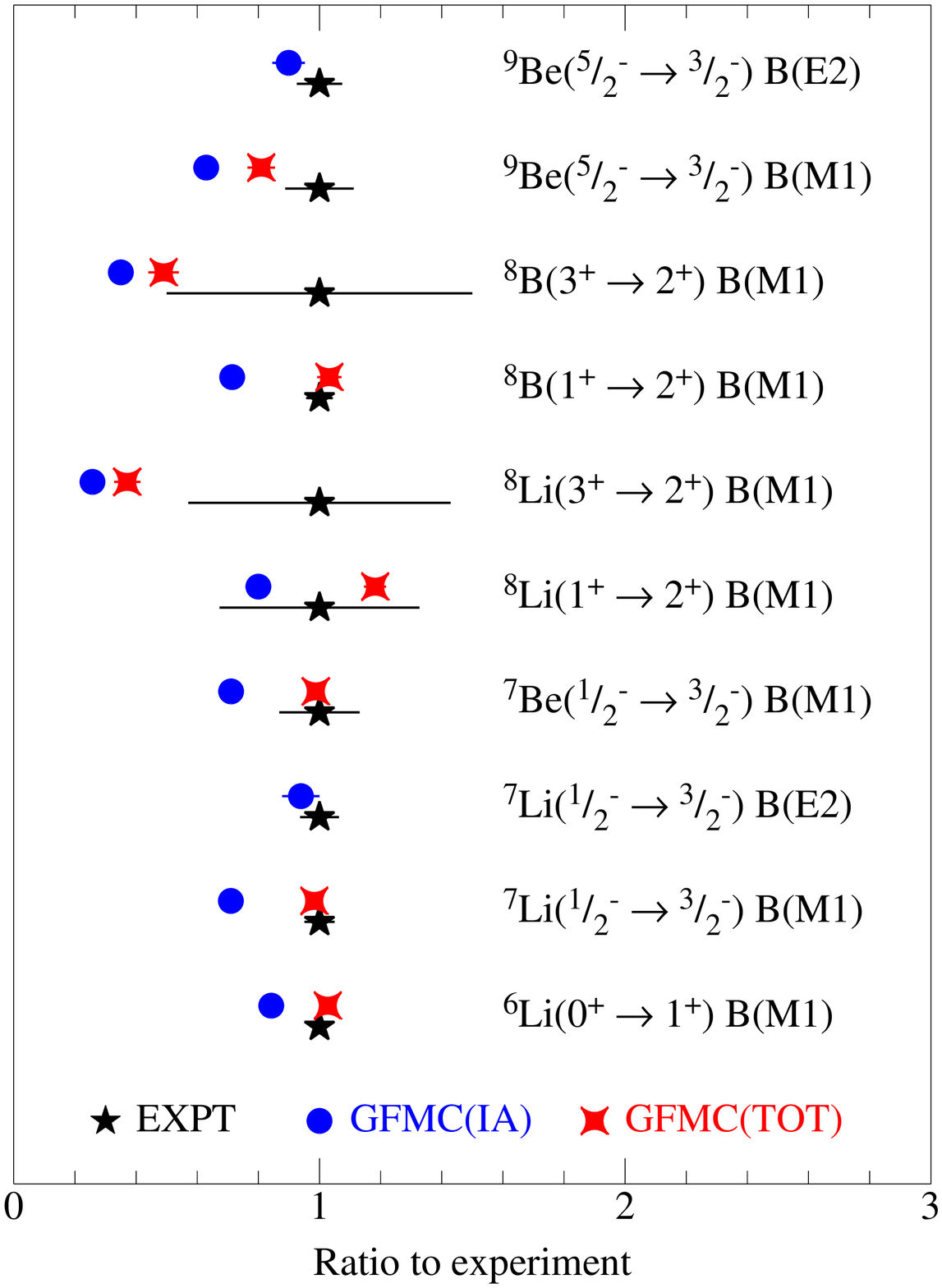}
\end{center}
\end{minipage}
\hspace{-0.5cm}
\begin{minipage}[t]{0.30\textheight}
\vspace{-2.5in}
\includegraphics[height=.35\textheight,angle=270,keepaspectratio=true]{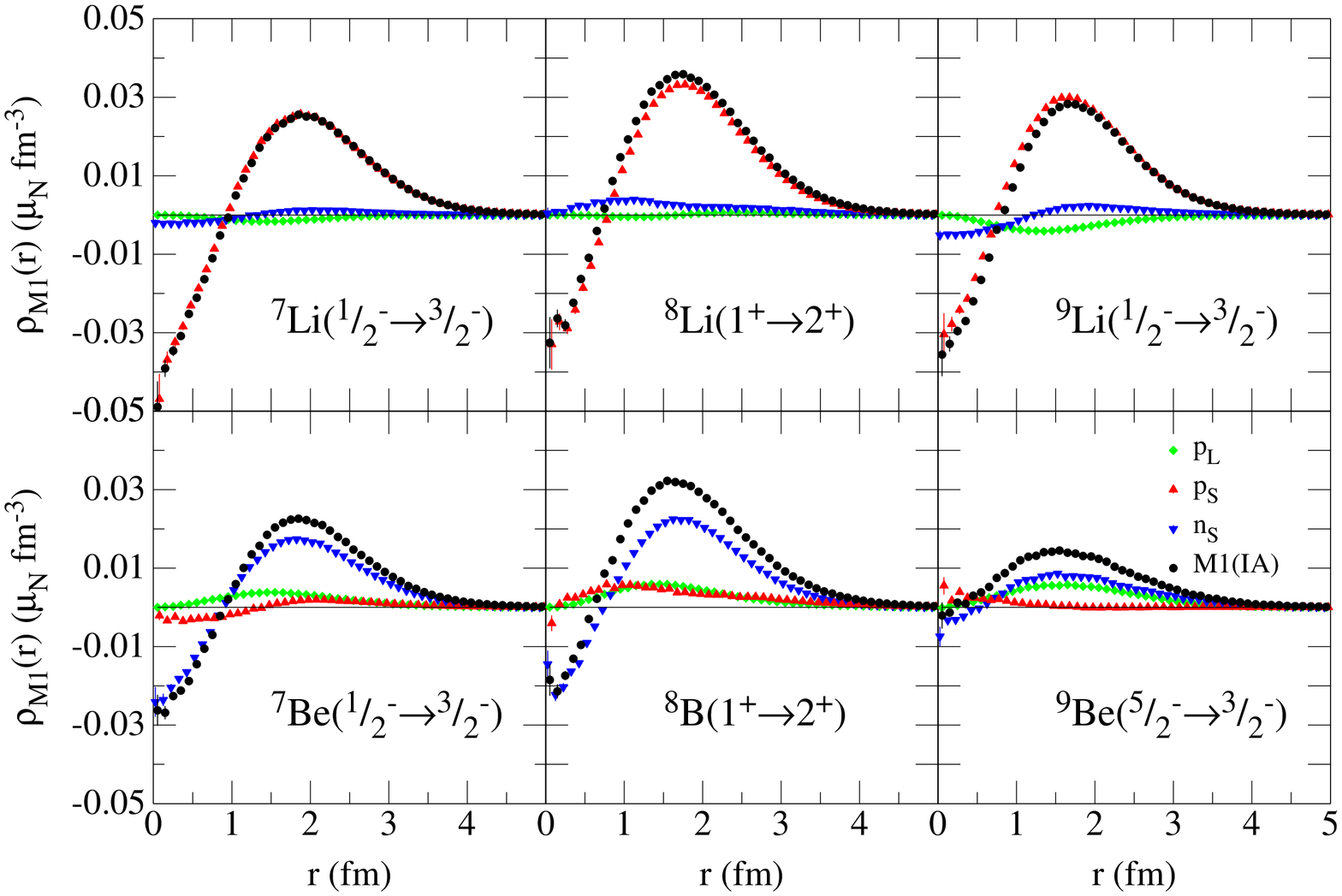}
\label{fig:f3}
\end{minipage}
\caption{  {\bf Left}: Ratio to the experimental $M1$ and $E2$ transition widths in $A\leq9$ nuclei.
           Black stars with error bars indicate the experimental 
           values~\cite{Tilley02,Tilley04}, while blue dots (red diamonds)
           represent GFMC calculations which include the IA one-body EM current
           (total $\chi$EFT current up to N3LO). {\bf Right}: $M1$ transition density in nuclear
           magnetons per fm$^3$  for selected nuclei, including only the IA current contribution.
           }
\end{figure}
Results for the m.m.'s of $A\leq 9$ nuclei indicate that two-body MEC corrections
evaluated in both the SNPA and $\chi$EFT models are qualitatively in agreement,
and, when large, they boost the IA in the direction of the experimental data.
We summarize the m.m.'s calculations  in the left panel of Fig.~\ref{fig:f2}, where
we show the results obtained with the $\chi$EFT model.
In this figure, black stars represent the experimental data~\cite{Tilley02,Tilley04}---there are no data
for the m.m. of $^9$B.\footnote{Electromagnetic moments of lithium isotopes have been most recently measured in
Refs.~\cite{Borremans05,Neugart08}. At the present time, the tables and figures of this contribution and
the preprint of Ref.~\cite{Pastore12} show, for these nuclei, the experimental data taken from
Refs.~\cite{Tilley02,Tilley04}.} For completeness, we show also the
experimental values for the proton and neutron m.m.'s, as well as their sum, which corresponds to the m.m.
of an S-wave deuteron. The experimental values of the $A=2$--$3$ m.m.'s have been utilized to fix the LECs,
therefore predictions are for $A>3$ nuclei. The blue dots labeled as GFMC(IA) represent IA theoretical
predictions. The GFMC(IA) results reproduce the bulk properties of the m.m.'s
of the light nuclei considered here. In particular, we can recognize three classes of nuclei, that is
nuclei whose m.m.'s  are driven by an unpaired valence proton, or neutron,
or `deuteron cluster' inside the nucleus. This behavior can be appreciated by looking at the
IA magnetic densities represented in the right panel of Fig.~\ref{fig:f2}, where
the red upward-pointing triangles are the contribution from
the proton spin, $\mu_p[\rho_{p\uparrow}(r) - \rho_{p\downarrow}(r)]$, 
and similarly the blue downward-pointing triangles are the contribution 
from the neutron spin. The green diamonds are the proton orbital (convection current)
contribution, and the black circles are the sum.
For example, we can see that the m.m.'s of $^7$Li and $^9$Li are driven by the unpaired proton,
while the m.m. of $^8$Li it is due to a combined effect of the unpaired neutron acting against the proton.

In the left panel of Fig.~\ref{fig:f2}, predictions, which include all the $\chi$EFT EM current
contributions illustrated in Fig.~\ref{fig:f1}, are represented by the red
diamonds labeled GFMC(TOT). In all of the cases considered
here---except for $^6$Li and $^9$Be for which the IA results are already in 
very good agreement with the experimental data, the predicted m.m.'s  are closer to the
experimental data when the MEC corrections are added to the IA results. 
MEC corrections are particularly pronounced in the isovector combination of the $A=9$, $T=3/2$
nuclei's m.m.'s, for which the MEC SNPA ($\chi$EFT) correction provides $\sim20\%$
($\sim 30\%$) of the total calculated isovector contribution. While the SNPA and $\chi$EFT
models are in a reasonable good agreement when predicting the isovector
m.m.'s---which are driven by the long-range NLO OPE contribution,
we find that isoscalar m.m.'s evaluated within the $\chi$EFT model are usually 
in a better agreement with the experimental data~\cite{Pastore12}.

In the left panel of Fig.~\ref{fig:f3}, we show EM transitions induced by the M1 and E2
operators in $A\leq 9$ nuclei---E2 transitions are provided in IA only.
In this figure, we show the ratios to the experimental values of the widths~\cite{Tilley02,Tilley04}.
The latter are represented with the black stars along with the associated experimental error bars,
while the GFMC(IA) and GFMC(TOT) predictions are again represented by blue dots and red diamonds,
respectively. For the M1 transition in IA, we also provide their transition densities which
are illustrated in the right panel of Fig.~\ref{fig:f3}. As before, the red upward-pointing
triangles are the contribution from the proton spin term, the blue downward-pointing triangles are from the neutron
spin, the green diamonds are from the proton orbital term, and the black 
circles are the total IA contribution. For example, for the lithium isotopes,
the M1 IA transitions are predominantly from the proton
spin term, {\it i.e.}, these are almost pure proton spin-flip transitions.
While, for $^7$Be and $^8$B, the neutron spin term is the most important, but with
some contribution from the proton spin and orbital terms.
The M1 results summarized in the left panel of Fig.~\ref{fig:f3},
indicate that, also for these observables, predictions which account for MEC corrections
are closer to the experimental values, but for the transition in $^8$Li, for which the experimental
error is large, we cannot determine whether the GFMC(TOT) prediction is a better one.

This work is supported by the U.S. Department
of Energy, Office of Nuclear Physics, under contracts
No. DE-AC02-06CH11357 and No. DE-AC05-06OR23177, under the NUCLEI
SciDAC-3 grant, and under NSF grant PHY-1068305.


\vspace*{-0.18in}
\begin{thebibliography}{99}
%
\bibitem{Pastore12}
S.\ Pastore, S. C.\ Pieper, R.\ Schiavilla, and R. B.\ Wiringa, arXiv:1212.3375.
%
\bibitem{WSS95}
R. B. Wiringa, V. G. J. Stoks, and R. Schiavilla,
Phys. Rev. C {\bf 51}, 38 (1995).
%
\bibitem{P08b}
S. C. Pieper,
AIP Conf. Proc. {\bf 1011}, 143 (2008).
%
\bibitem{Mar05} L.\ E.\ Marcucci, M.\ Viviani, R.\ Schiavilla, A.\ Kievsky,
                and S.\ Rosati,
                Phys.\ Rev.\ C {\bf 72}, 014001 (2005).
%
\bibitem{MPPSW08}
L. E. Marcucci, M. Pervin, S. C. Pieper, R. Schiavilla, and R. B. Wiringa,
Phys. Rev. C {\bf 78}, 065501 (2008).
%
\bibitem{Pastore08}
S.\ Pastore, R.\ Schiavilla, and J.L.\ Goity,
Phys.\ Rev.\ C {\bf 78}, 064002 (2008).
%
\bibitem{PGSVW09}
S. Pastore, L. Girlanda, R. Schiavilla, M. Viviani, and R. B. Wiringa,
Phys. Rev. C {\bf 80}, 034004 (2009).
%
\bibitem{Piarulli12}
M.\ Piarulli, L.\ Girlanda, L.E.\ Marcucci, S.\ Pastore, R.\ Schiavilla, and M. \ Viviani,
Phys. Rev. C {\bf 87}, 014006 (2013).
%
\bibitem{PPW07}
M. Pervin, S. C. Pieper, and R. B. Wiringa,
Phys. Rev. C {\bf 76}, 064319 (2007).

%
\bibitem{Park96}
T.-S.\ Park, D.-P.\ Min, and M.\ Rho,
Nucl.\ Phys.\ {\bf A596}, 515 (1996).
%
\bibitem{Kolling09}
S.\ K\"olling, E.\ Epelbaum, H.\ Krebs, U.-G.\ Meissner,
Phys.\ Rev.\ {\bf C80}, 045502 (2009).
%
\bibitem{Kolling11}
S.\ K\"olling, E.\ Epelbaum, H.\ Krebs, and U.-G.\ Meissner,
Phys.\ Rev.\ C {\bf 84}, 054008 (2011).
%
%
\bibitem{Song07}
Y.-H.\ Song, R.\ Lazauskas, T.-S.\ Park, and D.-P.\ Min
Phys. Lett. B 656, 174 (2007).
%
\bibitem{Song09}
Y.-H.\ Song, R.\ Lazauskas, and T.-S.\ Park,
Phys. Rev. C 79, 064002 (2009).
%
\bibitem{Lazauskas11}
R.\ Lazauskas, Y.-H.\ Song, and T.-S.\ Park,
Phys. Rev. C 83, 034006 (2011).
%
%
\bibitem{Entem03} 
D.R.\ Entem and R.\ Machleidt,
Phys.\ Rev.\ C {\bf 68}, 041001 (2003);
R.\ Machleidt and D.R.\ Entem,
Phys.\ Rep.\ {\bf 503}, 1 (2011).
%
%
\bibitem{Tilley02}
D. R. Tilley, C. M. Cheves, J. L. Godwin, G. M. Hale, H. M. Hofmann,
J. H. Kelley, C. G. Sheu, and H. R. Weller,
Nucl. Phys. A {\bf 708}, 3 (2002).
%
\bibitem{Tilley04}
D. R. Tilley, J. H. Kelley, J. L. Godwin,D. J. Millener, J. E. Purcell,
C. G. Sheu, and H. R. Weller,
Nucl. Phys. A {\bf 745}, 155 (2004).
%
\bibitem{Borremans05}
D.\ Borremans {\it et al.},
Phys.\ Rev.\ C {\bf 72}, 044309 (2005).
%
\bibitem{Neugart08}
R.\ Neugart {\it et al.},
Phys. Rev. Lett. 101, 132502 (2008).


%
%

\end{thebibliography}
\end{document}